\documentclass[english,aps,pra,superscriptaddress,twocolumn,address,showpacs,showacknowledgments]{revtex4}
\usepackage[T1]{fontenc}
\usepackage[latin9]{inputenc}
\usepackage{babel}
\usepackage{amsmath}
\usepackage{amssymb}
\usepackage{graphicx}
\usepackage{esint}
\usepackage[unicode=true,
 bookmarks=true,bookmarksnumbered=false,bookmarksopen=false,
 breaklinks=false,pdfborder={0 0 1},backref=false,colorlinks=false]
 {hyperref}
\hypersetup{pdftitle={Arbitrarily large steady-state bosonic squeezing via dissipation},
 pdfauthor={Andreas Kronwald, Florian Marquardt, and Aashish A. Clerk}}

\makeatletter
\@ifundefined{textcolor}{}
{%
 \definecolor{BLACK}{gray}{0}
 \definecolor{WHITE}{gray}{1}
 \definecolor{RED}{rgb}{1,0,0}
 \definecolor{GREEN}{rgb}{0,1,0}
 \definecolor{BLUE}{rgb}{0,0,1}
 \definecolor{CYAN}{cmyk}{1,0,0,0}
 \definecolor{MAGENTA}{cmyk}{0,1,0,0}
 \definecolor{YELLOW}{cmyk}{0,0,1,0}
 }

\usepackage{bbold}
\renewcommand\[{\begin{equation}}
\renewcommand\]{\end{equation}} 

\renewenvironment{eqnarray*}{\eqnarray}{\endeqnarray}

\makeatother

\begin{document}

\title{Arbitrarily large steady-state bosonic squeezing via dissipation}

\author{Andreas Kronwald}

\email{andreas.kronwald@physik.uni-erlangen.de}

\selectlanguage{english}%

\affiliation{Friedrich-Alexander-Universität Erlangen-Nürnberg, Staudtstr. 7,
D-91058 Erlangen, Germany}

\author{Florian Marquardt}

\affiliation{Friedrich-Alexander-Universität Erlangen-Nürnberg, Staudtstr. 7,
D-91058 Erlangen, Germany}

\affiliation{Max Planck Institute for the Science of Light, Günther-Scharowsky-Straße
1/Bau 24, D-91058 Erlangen, Germany}

\author{Aashish A. Clerk}

\affiliation{Department of Physics, McGill University, Montreal, Quebec, Canada
H3A 2T8 }

\begin{abstract}
We discuss how large amounts of steady-state quantum squeezing (beyond
3 dB) of a mechanical resonator can be obtained by driving an optomechanical
cavity with two control lasers with differing amplitudes. The scheme
does not rely on any explicit measurement or feedback, nor does it
simply involve a modulation of an optical spring constant. Instead,
it uses a dissipative mechanism with the driven cavity acting as an
engineered reservoir. It can equivalently be viewed as a coherent
feedback process, obtained by minimally perturbing the quantum non-demolition
measurement of a single mechanical quadrature. This shows that in
general the concepts of coherent feedback schemes and reservoir engineering
are closely related. We analyze how to optimize the scheme, how the
squeezing scales with system parameters, and how it may be directly
detected from the cavity output. Our scheme is extremely general,
and could also be implemented with, e.g., superconducting circuits.
\end{abstract}
\maketitle
\global\long\def\i{i}
\global\long\def\G{\mathcal{G}}
\global\long\def\a{\hat{a}}
\global\long\def\ad{\hat{a}^{\dagger}}
\global\long\def\d{\hat{d}}
\global\long\def\dd{\hat{d}^{\dagger}}
\global\long\def\b{\hat{b}}
\global\long\def\bd{\hat{b}^{\dagger}}
\global\long\def\be{\hat{\beta}}
\global\long\def\bed{\hat{\beta}^{\dagger}}
\global\long\def\C{\mathcal{C}}
\global\long\def\nth{n_{\text{th}}}
\global\long\def\neff{n_{\text{eff}}}
\global\long\def\eps{\varepsilon}
\global\long\def\x{\hat{x}}
\global\long\def\p{\hat{p}}
\global\long\def\Xone{\hat{X_{1}}}
\global\long\def\Xtwo{\hat{X_{2}}}

\textit{Introduction \textendash{}} Among the simplest nonclassical
states of a harmonic oscillator are quantum squeezed states, where
the uncertainty of a single motional quadrature is suppressed below
the zero-point level~%
\footnote{We use the standard quantum optics definition where {}``nonclassical''
describes any state that cannot be described by a non-singular $P$-function
(see, e.g., \cite{2005_GerryKnight_IntroQuantumOptics}).%
}. Such states are of interest for a variety of applications in ultra-sensitive
force detection \cite{1980_Caves_ForceMsmntQuantumMechOsc}; they
are also a general resource for continuous variable quantum information
processing \cite{2005_BraunsteinRMP}. It has long been known that
coherent parametric driving can be used to generate squeezing of a
bosonic mode; for a mechanical resonator, this simply amounts to modulating
the spring constant at twice the mechanical resonance frequency \cite{2008_WallsMilburnBook}.
Such a simple parametric interaction can yield at best steady-state
squeezing by a factor of $1/2$ below the zero-point level (the so-called
3 dB limit) \cite{1981_MilburnWalls_3dBLimitOfSqueezing}; if one
further increases the interaction strength, the system becomes unstable
and starts to self-oscillate. 

The rapid progress in quantum optomechanics \cite{2008_Kippenberg_Mesoscale_Review,2009_FM_ReviewOptomechanics,2013_Meystre_OptomechanicsReview,2013_AspelmeyerKippenbergMarquardt_OptomechanicsReview},
where a driven electromagnetic cavity is used to detect and control
a mechanical resonator, has led to a renewed interest in the generation
of squeezing. One could simply use radiation pressure forces to define
an oscillating spring constant \cite{2009_MariEisert_GentlyModulating,2008_Woolley_NanomechanicalSqueezing,2010_Nunnenkamp_CoolingAndSqueezingQuadraticCoupling,2011_Liao_Law_ParametericSqueezingGeneration,2012_Schmidt_CVQuantumStateProcessing},
though this cannot surpass the usual 3 dB limit on stationary squeezing.
One can do better by combining continuous quantum measurements and
feedback, either by making a quantum non-demolition (QND) measurement
of a single motional quadrature \cite{1980_Braginsky_QND_Quadrature_Measurement,1980_Caves_ForceMsmntQuantumMechOsc,2005_Ruskov_SqueezingMechResByQND_Stroboscopic,2008_Clerk_BackActionEvasion},
or by combining detuned parametric driving with position measurement
\cite{2011_Szorkovszky_MechanicalSqueezingFeedback,2013_SzorkovszkyThermoSqueezing}.
While such schemes can generate quantum squeezing well past the 3
dB limit, they are difficult to implement, as they require near-ideal
measurements and feedback. The 3 dB limit could also be surpassed
by continuously injecting squeezed light directly into the cavity
\cite{2009_Jaehne_CavityAssistedSqueezingOfMechOscillator}, but this
is also difficult as one needs to start with a source of highly squeezed
light. We note that a mechanical resonator could also be squeezed
via a pulsed optomechanical scheme \cite{2011_Vanner_PulsedQuOptomechanics}.
\begin{figure}[b]
\includegraphics{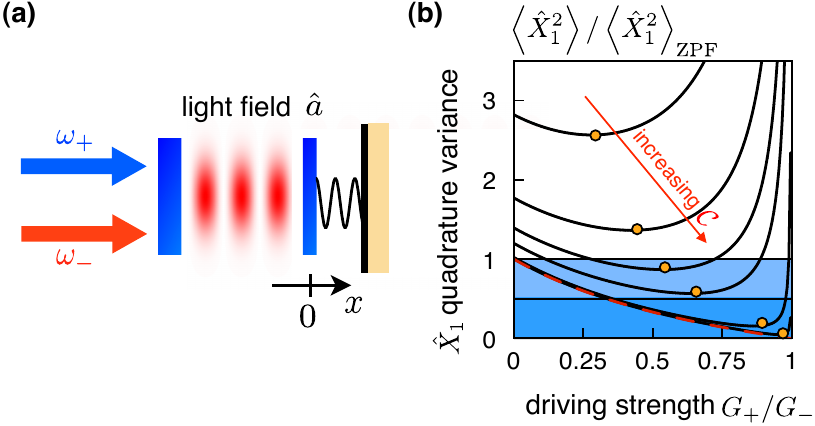}

\caption{(a) An optomechanical cavity is driven on the red and blue mechanical
sideband with different laser amplitudes. This leads to steady-state
mechanical squeezing beyond $3\,\mathrm{dB}$. (b) Steady-state quadrature
fluctuations $\left\langle \hat{X}_{1}^{2}\right\rangle $ {[}in units
of the zero point fluctuations (ZPF){]} as a function of the blue
laser driving strength $G_{+}$ for different cooperativities $\C=4G_{-}^{2}/\left(\kappa\Gamma_{M}\right)$.
The dark blue region indicates squeezing beyond $3\,\mathrm{dB}$.
The red, dashed line is the variance of a squeezed vacuum state with
squeeze parameter $r={\rm arctanh}\, G_{-}/G_{+}$. An optimal choice
of $G_{+}/G_{-}$ exists {[}orange circles{]} maximizing the amount
of squeezing for each $\C$. {[}Parameters: $\Gamma_{M}/\kappa=10^{-4}$,
$\nth=10$, $\C=10,25,50,10^{2},10^{3},10^{4}${]}.}

\label{fig:Setup}
\end{figure}

In this work, we discuss a remarkably simple scheme for generating
\emph{steady-state }mechanical squeezing well beyond the 3 dB level.
We use a two-tone driving of an optomechanical cavity, without any
explicit measurement or feedback (cf.~Fig.~\ref{fig:Setup}(a)).
As described in Refs. \cite{1980_Braginsky_QND_Quadrature_Measurement,1980_Caves_ForceMsmntQuantumMechOsc,2008_Clerk_BackActionEvasion},
if one drives the cavity with equal amplitude at both $\omega_{{\rm cav}}\pm\Omega$
(where $\omega_{\text{cav}}$ is the cavity frequency, $\Omega$ the
mechanical frequency), then the cavity only couples to a single mechanical
quadrature $X_{1}$, allowing for a QND measurement and preventing
any backaction disturbance of $X_{1}$. In contrast, we consider a
situation where the two drive tones have different amplitudes. One
thus no longer has a QND situation, and there will be a backaction
disturbance of $X_{1}$. However, this disturbance acts to \emph{suppress}
the fluctuations of $X_{1}$, to a level even below the zero-point
level. Our scheme thus realizes a coherent feedback operation, where
the driven cavity both measures the $X_{1}$ quadrature \emph{and
}autonomously applies the corresponding feedback operation necessary
to {}``cool'' $X_{1}$; (cf. appendix \ref{sec:Appendix:-Semiclassical-picture}
and Refs. \cite{PhysRevA.85.013812,2012_HamerlyMabuchi_CoherentFeedback,2013_KerckhoffLehnertCoherentFeedback}
for other examples of coherent feedback in optomechanics). This, hence,
suggests the general recipe to construct a coherent feedback scheme
by perturbing a QND measurement setup minimally. Equivalently, one
can think of the scheme as an example of reservoir engineering \cite{1996_PoyatosZollerReservoirEngineering}:
the driven cavity acts effectively as a bath whose force noise is
squeezed. This suggests that in general the concepts of coherent feedback
and reservoir engineering are closely related.

In what follows, we provide a thorough analysis of the optimal steady-state
squeezing generated by our scheme, showing that the squeezing is a
sensitive function of the ratio of the cavity drive amplitudes. We
also show that for realistic parameters, one can obtain mechanical
squeezing well beyond the usual 3 dB limit associated with a coherent
parametric driving. While we focus here on optomechanics, our scheme
could also be realized in other implementations of parametrically
coupled bosonic modes, e.g. superconducting circuits \cite{2010_Bergeal_JosephsonRingModulator,2013_Peropadre_TunableCouplingSupercondResonators}.
Note that related dissipative mechanisms can be utilized to prepare
the motion of a trapped ion in a squeezed state \cite{1993_Cirac_DarkSqueezedStates},
to squeeze a mechanical resonator which is coupled to a two-level-system
\cite{2004_Rabl_NanomechSqueezedStates} and to produce spin squeezing
of atoms in a cavity \cite{2006_Parkins_DissipativeTwoModeSqueezing,2013_DallaTorre_SteadyStateSqueezingSpinEnsemble}.
Unlike those works, our analysis does not rely on describing the engineered
reservoir (the driven cavity) via a simple Lindblad master equation;
in fact, we explicitly discuss corrections to such an approximation,
which we show to become significant for current experiments. We also
note that reservoir engineering approaches to optomechanics have been
previously considered for generating \textit{entanglement} (two-mode
squeezing) \cite{2013_WangClerk_SteadyStateEntanglement,2013_TanMeystreDissipativeEntanglement},
as well as coherence in arrays \cite{2012_TomadinReservoirEngineeringOMArrays}. 

\textit{Model \textendash{} }We consider a standard optomechanical
system, where a single cavity mode couples to a mechanical resonator
via radiation pressure, cf. Fig.~\ref{fig:Setup}(a). It is described
by the optomechanical Hamiltonian \cite{1995_Law_OptomechHamiltonian}
\begin{equation}
\hat{H}=\hbar\omega_{\text{cav}}\hat{a}^{\dagger}\hat{a}+\hbar\Omega\hat{b}^{\dagger}\hat{b}-\hbar g_{0}\hat{a}^{\dagger}\hat{a}\left(\hat{b}^{\dagger}+\hat{b}\right)+\hat{H}_{\text{dr}}\label{eq:OM_Ham}
\end{equation}
where the two-tone laser driving Hamiltonian reads
\[
\hat{H}_{\text{dr}}=\hbar\left(\alpha_{+}e^{-\i\omega_{+}t}+\alpha_{-}e^{-\i\omega_{-}t}\right)\hat{a}^{\dagger}+\text{h.c.}\,.
\]
$\a$ $(\b)$ is the photon (phonon) annihilation operator, and $g_{0}$
is the optomechanical coupling. $\omega_{\pm}$ and $\alpha_{\pm}$
are the frequency and amplitude of the two lasers, respectively. We
apply the displacement transformation $\a=\bar{a}_{+}e^{-\i\omega_{+}t}+\bar{a}_{-}e^{-\i\omega_{-}t}+\d$
to (\ref{eq:OM_Ham}) and go into an interaction picture with respect
to the free cavity and mechanical resonator Hamiltonian. Here, $\bar{a}_{\pm}$
is the coherent light field amplitude due to the two lasers. If $\Omega$
is strongly temperature dependent, oscillations in the average cavity
intensity can yield spurious parametric instabilities \cite{2012_SuhBAEParampInstability};
these could be suppressed by adding an appropriate third drive tone
without strongly degrading the generation of squeezing \cite{2013_Steinke_BackacktionEvadingMsrmt}
(cf. appendix \ref{sec:Appendix:-Avoiding-the}). 

We next take the two lasers to drive the mechanical sidebands of a
common mean frequency $\bar{\omega}$, i.e. $\omega_{\pm}=\bar{\omega}\pm\Omega$,
and assume $|\bar{a}_{\pm}|\gg1$. For $\bar{\omega}$ far detuned
from $\omega_{{\rm cav}}$, one can eliminate the cavity to obtain
an {}``optical spring'' which is modulated at $2\Omega$ \cite{2011_Liao_Law_ParametericSqueezingGeneration,2012_Schmidt_CVQuantumStateProcessing};
Ref.~\cite{2009_MariEisert_GentlyModulating} obtains an analogous
effect by weakly amplitude-modulating a single strong drive at $\omega_{{\rm cav}}-\Omega$.
In contrast, we take $\bar{\omega}=\omega_{\text{cav}}$ as well as
$\bar{a}_{+}\neq\bar{a}_{-}$ (in contrast to back-action evasion
(BAE) schemes \cite{1980_Braginsky_QND_Quadrature_Measurement,1980_Caves_ForceMsmntQuantumMechOsc,2008_Clerk_BackActionEvasion}).
Applying a standard linearization to Eq.~(\ref{eq:OM_Ham}), we find
that the \textit{linearized} Hamiltonian in our interaction picture
is 
\begin{eqnarray}
\hat{H} & = & -\hbar\hat{d}^{\dagger}\left(G_{+}\hat{b}^{\dagger}+G_{-}\hat{b}\right)+\text{h.c.}\nonumber \\
 &  & -\hbar\hat{d}^{\dagger}\left(G_{+}\hat{b}e^{-2\i\Omega t}+G_{-}\hat{b}^{\dagger}e^{2\i\Omega t}\right)+\text{h.c.}.\label{eq:linearized_Hamiltonian}
\end{eqnarray}
Here, $G_{\pm}=g_{0}\bar{a}_{\pm}$ are the enhanced optomechanical
coupling rates; without loss of generality, we assume $G_{+},G_{-}>0$.
The quantum Langevin equations describing the dissipative dynamics
read \cite{2008_ClerkDevoretGirvinFMSchoelkopf_RMP}
\[
\dot{\d}=\frac{\i}{\hbar}\left[\hat{H},\d\right]-\frac{\kappa}{2}\d+\sqrt{\kappa}\d_{\text{in}}\,.
\]
A similar equation holds for $\b$, where the cavity decay rate $\kappa$
is replaced by the mechanical decay rate $\Gamma_{M}$. The non-zero
input noise correlators read $\langle\d_{\text{in}}\left(t\right)\hat{d}_{\text{in}}^{\dagger}\left(t'\right)\rangle=\delta\left(t-t'\right)$,
$\langle\b_{\text{in}}\left(t\right)\bd_{\text{in}}\left(t'\right)\rangle=(\nth+1)\delta\left(t-t'\right)$
and $\langle\bd_{\text{in}}\left(t\right)\b_{\text{in}}\left(t'\right)\rangle=\nth\delta\left(t-t'\right)$,
where $\nth$ is the thermal occupancy of the mechanical bath.

\textit{Squeezing generation \textendash{} }We now present two intuitive
ways of understanding the generation of steady-state squeezing in
our scheme. For physical transparency, we focus on the good cavity
limit $\kappa\ll\Omega$, and thus ignore counter-rotating terms in
Eq.~(\ref{eq:linearized_Hamiltonian}) until the last section.

Note first that if $G_{+}=G_{-}$ (i.e.~equal drive amplitudes),
the cavity only couples to the mechanical quadrature $\hat{X}_{1}=\left(\bd+\b\right)/\sqrt{2}$,
cf. Eq. (\ref{eq:linearized_Hamiltonian}). As discussed earlier,
this allows a QND measurement of $\hat{X}_{1}$ \cite{1980_Braginsky_QND_Quadrature_Measurement,2008_Clerk_BackActionEvasion}.
If $G_{+}\not=G_{-}$, the cavity still couples to a single mechanical
operator, a Bogoliubov-mode annihilation operator 
\[
\be=\b\cosh r+\bd\sinh r\,,
\]
where the squeezing parameter $r$ is defined via $\tanh r=G_{+}/G_{-}$.
We also assume $G_{+}<G_{-}$ which ensures stability. The Hamiltonian
(\ref{eq:linearized_Hamiltonian}) becomes
\begin{equation}
\hat{H}=-\G\dd\be+\text{h.c.}\label{eq:H_Bogol_mode}
\end{equation}
where the coupling $\G=\sqrt{G_{-}^{2}-G_{+}^{2}}$. This is a beam-splitter
Hamiltonian well known from optomechanical sideband cooling \cite{2007_FM_SidebandCooling,2007_Wilson-Rae_TheoryGroundStateCooling}.
However, instead of allowing the cavity to cool the mechanical mode,
it now can cool the mode $\be$. As the vacuum of $\be$ is the squeezed
state $\hat{S}\left(r\right)\left|0\right\rangle $ (where $\hat{S}\left(r\right)=\exp[r(\b\b+\bd\bd)/2]$)
\cite{2008_WallsMilburnBook}, this cooling directly yields steady-state
squeezing. In general,
\begin{equation}
2\left\langle \hat{X}_{1}^{2}\right\rangle =e^{-2r}\left[1+2\left\langle \bed\be\right\rangle +\left\langle \be\be\right\rangle +\left\langle \bed\bed\right\rangle \right]\,.\label{eq:X1_var_Bogol_mode}
\end{equation}
\begin{figure}[t]
\includegraphics{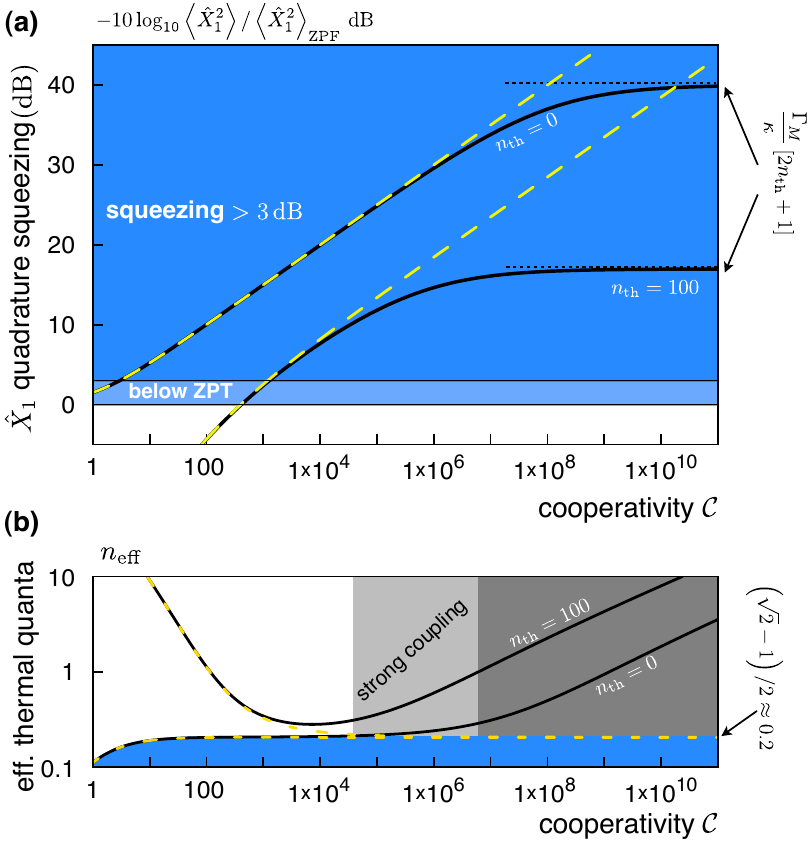}

\caption{Maximized steady-state squeezing and state purity. (a) Squeezing for
fixed $\Gamma_{M}/\kappa$ and optimized driving strength $G_{+}$
(cf. (\ref{eq:optim_driving_strength})), as a function of $G_{-}$
(parametrized by the cooperativity $\C$). Squeezing beyond $3\,\mathrm{dB}$
is apparent even for moderate $\C$. Black lines represent the full
theory. These curves are well described by (\ref{eq:min_variance_approx})
for $\C\gtrsim100$. Yellow, dashed lines show the prediction of a
Lindblad master equation (LME, Eq. (\ref{eq:eff_LME_mechanics}))
. (b) Effective thermal occupancy $\neff$ in the mechanical steady
state, for optimized parameters, as a function of $\C$. The colors
are the same as in (a). The full theory and the LME differ drastically
in the strong-coupling regime where $\G\gtrsim\kappa$ {[}gray shaded
region, beginning first for $\nth=100${]}. {[}Parameters: $\Gamma_{M}/\kappa=10^{-4}$,
$\nth=0$ and $\nth=100${]}.}

\label{fig:x1varmin_vs_coop_and_impurity}
\end{figure}
If $\be$ is in its groundstate, $2\left\langle \hat{X}_{1}^{2}\right\rangle =e^{-2r}$.
Thus, the cavity acts as an engineered reservoir that can cool the
mechanical resonator into a squeezed state. We note that related entanglement-via-dissipation
schemes \cite{2006_Parkins_DissipativeTwoModeSqueezing,2013_TanMeystreDissipativeEntanglement,2013_WangClerk_SteadyStateEntanglement}
are based on cooling \textit{a delocalized} Bogoliubov mode. In contrast,
we study a \textit{localized} mode which directly leads to a (single-mode)
squeezed mechanical steady-state.

As mentioned, one can also interpret the squeezing generation without
invoking a Bogoliubov mode, but rather as a coherent feedback operation
where the cavity both measures and perturbs $\Xone$. In the simplest
large-$\kappa$ limit, the feedback causes both $\Xone$ and $\Xtwo=\i\left(\bd-\b\right)/\sqrt{2}$
to be damped at a rate $\Gamma_{\text{opt}}=4\G^{2}/\kappa$, but
adds negligible fluctuations to $\Xone$ (smaller than the zero-point
fluctuations that one would associate with $\Gamma_{\text{opt }},$
cf. appendix \ref{sec:Appendix:-Semiclassical-picture}). Thus, despite
being driven with classical light, the cavity acts as a squeezed reservoir
leading to mechanical squeezing. 

\textit{Squeezing versus driving strengths \textendash{} }We solve
the quantum Langevin equations (first in the rotating-wave approximation)
and consider the steady-state mechanical squeezing as a function of
$G_{+}/G_{-}$, cf.~Fig.~\ref{fig:Setup}(b), holding constant both
the ratio of damping rates $\Gamma_{M}/\kappa$ and the red-laser
amplitude (parameterized via the cooperativity $\C=4G_{-}^{2}/\left(\kappa\Gamma_{M}\right)$).
Without the blue-detuned laser, i.e. $G_{+}=0$, we have standard
optomechanical sideband-cooling: both quadrature variances are reduced
as compared to the thermal case \cite{2007_FM_SidebandCooling,2007_Wilson-Rae_TheoryGroundStateCooling}.
Turning on $G_{+}$, the quadrature variance $\left\langle \hat{X}_{1}^{2}\right\rangle $
first decreases with increasing $G_{+}/G_{-}$. In general, $\left\langle \hat{X}_{1}^{2}\right\rangle $
exhibits a minimum as a function of $G_{+}/G_{-}$ which becomes sharper
with increasing cooperativity $\C$. For large $\C$, the minimum
variance is well below $1/2$ the zero-point value, i.e. the $3\,\mathrm{dB}$
limit. This minimum results from the competition of two opposing tendencies.
On one hand, increasing $G_{+}/G_{-}$ increases the squeezing parameter
$r$ and thus the squeezing associated with the vacuum of $\beta$.
On the other hand, increasing $G_{+}/G_{-}$ reduces $\G$, and hence
suppresses the ability of the cavity to cool $\beta$. The optimum
squeezing is thus a tradeoff between these tendencies.

\textit{Optimal squeezing \textendash{}} Consider a fixed red-laser
amplitude (i.e. $G_{-})$ large enough that the cooperativity $\C\gg1.$
The value of $G_{+}$ which maximizes squeezing is then:

\begin{equation}
\frac{G_{+}}{G_{-}}\Big|_{{\rm optimal}}\approx1-\sqrt{\frac{1+2n_{\text{th}}}{\C}}\,,\text{ i.e. }e^{-2r}\approx\frac{1}{2}\sqrt{\frac{1+2n_{\text{th}}}{\C}}.\label{eq:optim_driving_strength}
\end{equation}
The corresponding minimum value of $\left\langle \hat{X}_{1}^{2}\right\rangle $
is 
\begin{equation}
2\left\langle \hat{X}_{1}^{2}\right\rangle \approx\frac{\Gamma_{M}}{\kappa}\left(1+2n_{\text{th}}\right)+\sqrt{\frac{1+2n_{\text{th}}}{\C}}\,,\label{eq:min_variance_approx}
\end{equation}
cf. Fig. \ref{fig:x1varmin_vs_coop_and_impurity}(a). We see that
even for moderate values of $\C$ and non-zero $n_{{\rm th}}$, quantum
squeezing beyond $3\,\mathrm{dB}$ is achieved, cf.~Fig.~\ref{fig:x1varmin_vs_coop_and_impurity}(a).
As $\C$ is increased further, the amount of squeezing saturates to
a level set by the ratio of the mechanical heating rate to $\kappa$.
Note that if one attempts to describe the effect of the cavity on
the mechanical resonator via an effective Lindblad master equation,
one misses this saturation, cf.~Fig.~\ref{fig:x1varmin_vs_coop_and_impurity}(a)
(see appendix \ref{sec:Appendix:-effective-Lindblad}). An analogous
Lindblad approach was recently analyzed in the context of spin squeezing
in Ref. \cite{2013_DallaTorre_SteadyStateSqueezingSpinEnsemble}. 

\textit{State purity \textendash{}} It follows from Eqs.~(\ref{eq:X1_var_Bogol_mode})
- (\ref{eq:min_variance_approx}) that the maximal squeezing of our
scheme (at fixed cooperativity $\C$) corresponds to $\langle\hat{\beta}^{\dagger}\be\rangle>0$.
The steady state is thus a squeezed thermal state. To quantify the
purity of this state, we define an effective thermal occupancy from
the determinant of the mechanical covariance matrix, i.e.
\begin{equation}
\left(1+2\neff\right)^{2}=4\left\langle \hat{X}_{1}^{2}\right\rangle \left\langle \hat{X}_{2}^{2}\right\rangle -4\left\langle \{\Xone,\Xtwo\}\right\rangle ^{2}\,.\label{eq:thermalness_squeezed_state-1}
\end{equation}
The mechanical state is a pure squeezed vacuum state if $\neff=0$,
while the mixedness of the state increases with $n_{{\rm eff}}$.
As shown in Fig.~\ref{fig:x1varmin_vs_coop_and_impurity}(b), for
moderately-strong $\C$, one can both achieve squeezing beyond $3\,\mathrm{dB}$
and a low-entropy state, with $n_{{\rm eff}}\sim\left(\sqrt{2}-1\right)/2\approx0.2$
(independent of all parameters, see appendix \ref{sec:Appendix:-Squeezed-State}).
This again is in marked contrast to coherent parametric driving, where
the maximal squeezing of $3\,\mathrm{dB}$ is associated with a diverging
$n_{{\rm eff}}$. This is also in contrast to the squeezing generated
by a BAE measurement and feedback, where strong squeezing is also
associated with $n_{{\rm eff}}\gg1$ (cf. \cite{2008_Clerk_BackActionEvasion}
and appendix \ref{sec:Appendix:-Comparison-against}). 

As we increase $\C$ further, we enter the strong coupling regime
where $\G\gtrsim\kappa,\Gamma_{M}$ and the cavity and the Bogoliubov
mode hybridize. The squeezing saturates in this regime (cf.\ Eq.\ (\ref{eq:min_variance_approx})),
whereas $n_{{\rm eff}}$ increases without bound. For optimized couplings
we find $n_{{\rm eff}}^{2}\sim(\Gamma_{M}/2\kappa)\sqrt{(1+2n_{{\rm th}})\C}$
in the large $\C$ limit. Thus, while one can get enhanced squeezing
in the strong coupling regime, it comes at the price of strongly reducing
the purity of the squeezed state. It is also worth emphasizing that,
as shown in Fig. \ref{fig:x1varmin_vs_coop_and_impurity}(b), the
Lindblad master equation approximation fails to describe accurately
both the quadrature squeezing and $n_{{\rm eff}}$ at strong coupling;
this is not surprising as the Lindblad approach cannot describe the
hybdridization physics important in this regime, cf. appendix \ref{sec:Appendix:-effective-Lindblad}. 

\begin{figure}[t]
\includegraphics[width=1\columnwidth]{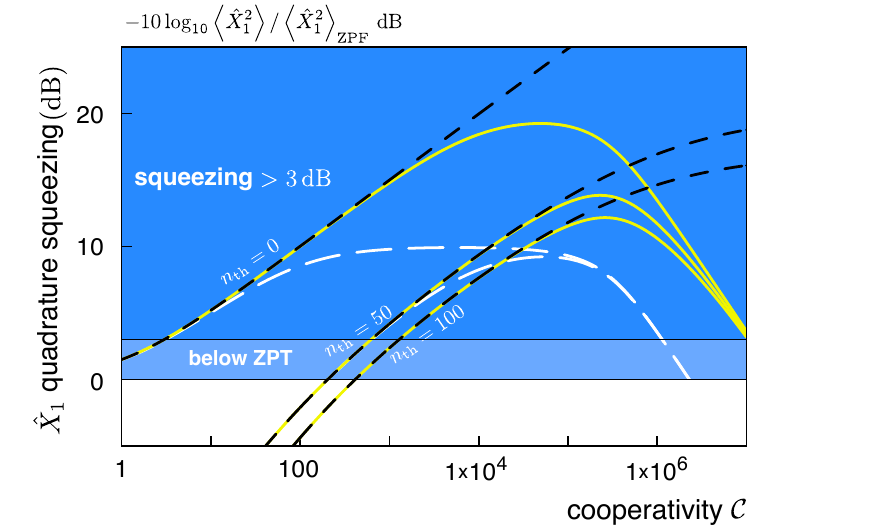}

\caption{Squeezing versus $\C$ for optimized driving strengths, using realistic
experimental parameters \cite{2011_Teufel_SidebandCooling_Nature},
and the full theory. Yellow and white dashed lines show the theory
\textit{including} effects due to a nonzero sideband parameter (i.e.~no
rotating-wave approximation, in contrast to Fig. \ref{fig:x1varmin_vs_coop_and_impurity}(a)).
The black, dashed line shows the expectation for $\kappa/\Omega=0$.
Thus, squeezing beyond $3\,\mathrm{dB}$ is expected for state-of-the-art
experiments. {[}Parameters: Yellow lines: $\Gamma_{M}/\kappa=10^{-4}$,
$\kappa/\Omega=1/50$. White dashed lines: $\Gamma_{M}/\kappa=10^{-5},$
$\kappa/\Omega=1/5${]}.}

\label{fig:influence_cr_terms}
\end{figure}

\textit{Squeezing detection\textendash{} }Squeezing of $X_{1}$ can
be detected by making a single-quadrature back-action evading measurement.
One generates the squeezed state as described above by having $G_{+}<G_{-}$.
To then measure the squeezing, one simply increases $G_{+}$ so that
$G_{+}=G_{-}$, thus allowing a QND measurement of $X_{1}$ \cite{1980_Braginsky_QND_Quadrature_Measurement,2008_Clerk_BackActionEvasion,2013_Steinke_BackacktionEvadingMsrmt}.
The measurement must be fast compared to the rate at which the mechanical
dissipation re-thermalizes $X_{1}$, leading to the condition $\C\gtrsim2\nth+1$
\cite{2008_Clerk_BackActionEvasion}. 

A simpler method for verifying squeezing is to keep $G_{+}<G_{-}$
and use the cavity output spectrum to extract the occupancy of the
$\hat{\beta}$ mode: photons in the output are a measure of how {}``hot''
this mode is. This is analogous to how mechanical temperature is obtained
in sideband cooling setups \cite{2007_FM_SidebandCooling,2007_Wilson-Rae_TheoryGroundStateCooling}.
The output spectrum is given by 
\[
S\left[\omega\right]=\int\mathrm{d}t\, e^{\i\omega t}\left\langle \delta\hat{a}_{\text{out}}^{\dagger}\left(t\right)\delta\hat{a}_{\text{out}}(0)\right\rangle \,.
\]
Here, $\delta\hat{a}_{\text{out}}=\hat{a}_{\text{out}}-\left\langle \hat{a}_{\text{out}}\right\rangle $,
and $\hat{a}_{\text{out}}+\hat{a}_{\text{in}}=\sqrt{\kappa}\hat{a}$
\cite{2008_ClerkDevoretGirvinFMSchoelkopf_RMP}, where we have assumed
an ideal, single-sided cavity. We find in the good cavity limit (cf.
appendix \ref{sec:Appendix:-Cavity-output}): 
\begin{equation}
\int\mathrm{d\omega}\, S[\omega]=8\pi\kappa\frac{\G^{2}}{4\G^{2}+\kappa\left(\kappa+\Gamma_{M}\right)}\left\langle \bed\be\right\rangle \,.\label{eq:area_s_sim_bogol_occ}
\end{equation}
Thus, knowing the red and blue driving strength $G_{\pm}$ as well
as the cavity decay rate $\kappa\gg\Gamma_{M}$ is sufficient to measure
$\left\langle \bed\be\right\rangle $. The knowledge of $\left\langle \bed\be\right\rangle $
is enough to find a \textit{rigorous} upper bound of the squeezing,
since
\[
\left\langle \hat{X}_{1}^{2}\right\rangle \le e^{-2r}\left[1+2\left\langle \bed\be\right\rangle \right]\,,
\]
(see appendix \ref{sec:Appendix:-Cavity-output}). For large $\C$,
this upper bound coincides with the actual value of $\langle\hat{X}_{1}^{2}\rangle$
(up to corrections $\sim1/\sqrt{\C}$). 

\textit{Effects of counter-rotating terms \textendash{} }\textit{\emph{We
now turn to the effects of the counter-rotating terms in Eq.}}~(\ref{eq:linearized_Hamiltonian})
which can play a role when one deviates from the extreme good cavity
limit $\kappa\ll\Omega$. These additional terms cause the cavity
to non-resonantly heat the Bogoliubov mode $\hat{\beta}$; as a result,
the coupling-optimized quadrature squeezing becomes a non-monotonic
function of the cooperativity $\C$. This is shown in Fig.~\ref{fig:influence_cr_terms},
where we have solved the full quantum Langevin equations, and taken
parameters from a recent experiment in microwave cavity optomechanics
\cite{2011_Teufel_SidebandCooling_Nature}. Steady-state squeezing
beyond $3\,\mathrm{dB}$ still exists even for only moderately resolved
sidebands ($\Omega/\kappa\sim5).$ To estimate the onset of the non-resonant
heating, we can calculate the leading $\mathcal{O}\left[\left(\kappa/\Omega\right)^{2}\right]$
correction to $\left\langle \hat{X}_{1}^{2}\right\rangle $. Insisting
it to be much smaller than the smallest variance possible in the extreme
good cavity limit (and taking $\C\gg1$) leads to the condition on
the cooperativity
\begin{equation}
\C^{3/2}\ll\sqrt{1+2\nth}\frac{\kappa}{\Gamma_{M}}\left(\frac{\Omega}{\kappa}\right)^{2}\,.\label{eq:CR_terms_negl_influence}
\end{equation}
This condition also ensures that the previous results for the optimized
coupling strengths remain valid. Further discussion of bad-cavity
effects (as well as parameters relevant to recent optical-frequency
optomechanics experiments) are presented in the appendix \ref{sec:Appendix:-Effects-of}.

\textit{Conclusion \textendash{} }We have shown that large steady-state
squeezing of a mechanical resonator can be achieved by driving an
optomechanical cavity at both the red and blue mechanical sideband,
with different amplitudes. For realistic parameters, steady-state
quantum squeezing well beyond the $3\,\mathrm{dB}$ limit can be generated.
By adding a final state transfer pulse, our scheme could also be used
to generate strong optical squeezing. It is also general enough to
be realized in other implementations of parametrically coupled bosonic
modes (e.g.~superconducting circuits).

\textit{Acknowledgements \textendash{}} We thank N. Didier and A.
Blais for drawing our attention to the possibility of dissipative
squeezing, and acknowledge support from the DARPA ORCHID program through
a grant from AFOSR, the Emmy-Noether program, the European Research
Council and the ITN cQOM. AK thanks AAC for his hospitality at McGill. 

Note added: A very recent preprint by Didier, Qassemi, and Blais \cite{2013_Didier_Squeezing}
analyzes an alternative dissipative squeezing mechanism.

\section{Appendix: Semiclassical picture of squeezing generation\label{sec:Appendix:-Semiclassical-picture}}

As described in the main text, one can obtain a semiclassical understanding
of the squeezing generation of our scheme by formally eliminating
the cavity from the dynamics; we provide more details on this approach
here. In addition to the mechanical resonator quadratures $\hat{X}_{1}$
and $\hat{X}_{2}$, we introduce the cavity quadratures by
\[
\hat{U}_{1}=\left(\dd+\d\right)/\sqrt{2}\,\text{ and }\,\hat{U}_{2}=\i\left(\dd-\d\right)/\sqrt{2}.
\]
Using the Hamiltonian (3), the Heisenberg-Langevin equations take
the form: 
\begin{align}
\dot{\hat{X}}_{1} & =-\left(G_{-}-G_{+}\right)\hat{U}_{2}-\frac{\Gamma_{M}}{2}\hat{X}_{1}+\sqrt{\Gamma_{M}}\hat{X}_{1,\text{in}}\nonumber \\
\dot{\hat{U}}_{2} & =\left(G_{-}+G_{+}\right)\hat{X}_{1}-\frac{\kappa}{2}\hat{U}_{2}+\sqrt{\kappa}\hat{U}_{2,\text{in}}\label{eq:EOM_coherent_feedback_1}
\end{align}
and
\begin{align}
\dot{\hat{U}}_{1} & =-\left(G_{-}-G_{+}\right)\hat{X}_{2}-\frac{\kappa}{2}\hat{U}_{1}+\sqrt{\kappa}\hat{U}_{1,\text{in}}\nonumber \\
\dot{\hat{X}}_{2} & =\left(G_{-}+G_{+}\right)\hat{U}_{1}-\frac{\Gamma_{M}}{2}\hat{X}_{2}+\sqrt{\Gamma_{M}}\hat{X}_{2,\text{in}}\label{eq:EOM_coherent_feedback_2}
\end{align}
where we have also introduced quadratures of the input noise operators
$\hat{d}_{{\rm in}}$ and $\hat{b}_{{\rm in}}$. Note that these are
two decoupled sets of equations. They immediately let us understand
the backaction-evading limit $G_{+}=G_{-}$ \cite{1980_Braginsky_QND_Quadrature_Measurement,2008_Clerk_BackActionEvasion}:
the cavity $\hat{U}_{2}$ quadrature measures the mechanical resonator's
quadrature $\hat{X}_{1}$ without disturbing its time-evolution. The
generation of steady-state squeezing, however, requires that $G_{+}$
be slightly smaller than $G_{-}$. In this case, the cavity $\hat{U}_{2}$
quadrature still measures $\hat{X}_{1}$. However, the cavity $\hat{U}_{2}$
quadrature also acts as a force on $\hat{X}_{1}$; we can view this
as a weak, coherent feedback force \cite{2012_HamerlyMabuchi_CoherentFeedback}
{[}cf. first line of Eq. (\ref{eq:EOM_coherent_feedback_1}){]}. As
we now show, this effective feedback directly leads to steady-state
squeezing. 

It is convenient to work in the Fourier domain, where
\[
f[\omega]=\frac{1}{\sqrt{2\pi}}\int\mathrm{d}t\, e^{\i\omega t}f(t)\,.
\]
 Eliminating the cavity quadratures from the mechanical equations
of motion, we find
\begin{align}
\left[-i\omega+\frac{\Gamma_{M}}{2}+\i\Sigma\left[\omega\right]\right]\hat{X}_{1}\left[\omega\right] & =-\sqrt{\kappa}\frac{i\Sigma[\omega]}{G_{-}+G_{+}}\,\hat{U}_{2,\text{in}}\left[\omega\right]\nonumber \\
 & +\sqrt{\Gamma_{M}}\hat{X}_{1,\text{in}}\left[\omega\right]
\end{align}
and
\begin{align}
\left[-i\omega+\frac{\Gamma_{M}}{2}+\i\Sigma\left[\omega\right]\right]\hat{X}_{2}\left[\omega\right] & =\sqrt{\kappa}\frac{i\Sigma[\omega]}{G_{-}-G_{+}}\,\,\hat{U}_{1,\text{in}}\left[\omega\right]\nonumber \\
 & +\sqrt{\Gamma_{M}}\hat{X}_{2,\text{in}}\left[\omega\right]\,.
\end{align}
where the self energy $\Sigma\left[\omega\right]=-i\left(G_{-}^{2}-G_{+}^{2}\right)/\left(\kappa/2-\i\omega\right)$
is the same for both quadratures. The imaginary part of $\Sigma$
describes damping of the mechanical quadratures by the cavity. These
equations also imply that the correlations $\left\langle \hat{X}_{1}\hat{X}_{2}+\hat{X}_{2}\hat{X}_{1}\right\rangle $
are zero. 

The cavity also introduces new noise terms driving each mechanical
quadrature. We can parameterize them by an effective temperature in
the standard way, by considering their magnitude compared to the corresponding
cavity-induced damping:
\begin{eqnarray*}
1+2n_{\text{eff},X_{1}}[\omega] & \equiv & \frac{\kappa|\Sigma[\omega]/(G_{-}+G_{+})|^{2}}{-2\,{\rm Im}\Sigma[\omega]}\,,\\
1+2n_{\text{eff},X_{2}}[\omega] & \equiv & \frac{\kappa|\Sigma[\omega]/(G_{-}-G_{+})|^{2}}{-2{\rm \, Im}\Sigma[\omega]}\,.
\end{eqnarray*}
Taking the low frequency limit, we have: 
\begin{eqnarray}
1+2n_{\text{eff},X_{1}}[0] & = & \frac{G_{-}-G_{+}}{G_{-}+G_{+}}\,,\nonumber \\
1+2n_{\text{eff},X_{2}}[0] & = & \frac{G_{-}+G_{+}}{G_{-}-G_{+}}\,.\label{eq:eff_noise_leading_to_squeezing-1}
\end{eqnarray}
Thus, while the cavity damps both mechanical quadratures the same
way, the noise added to the $\hat{X}_{1}$ quadrature is much smaller
than the noise added to the $\hat{X}_{2}$ quadrature (this is different
from coherent parametric driving, where the squeezed $\Xone$ quadrature
experiences extra damping whereas $\Xtwo$ experiences extra \textit{negative}
damping). Moreover, the magnitude of the noise added to the $\hat{X}_{1}$
quadrature is \emph{smaller }than the the zero-point noise one would
associate with the optical damping $\Gamma_{{\rm opt}}\equiv-2{\rm Im}\Sigma[0]$,
i.e.~$n_{{\rm eff,}X_{1}}[0]<0$. We thus see that the cavity effectively
acts as a squeezed reservoir, i.e. a reservoir whose force noise is
quadrature squeezed. If this cavity-induced dissipation dominates
the intrinsic mechanical dissipation, this directly yields squeezing
of the mechanical resonator. 

Finally, it is interesting to note that for a fixed $\C\gg1$, the
optimal ratio of $G_{+}/G_{-}$ given in Eq.~(8) of the main text
can be given a simple interpretation in terms of the effective optical
damping $\Gamma_{{\rm opt}}=4\mathcal{G}^{2}/\kappa$ introduced above
(with $\G^{2}=G_{-}^{2}-G_{+}^{2}$). Using the result of Eq.~(8)
we have:
\begin{eqnarray}
\Gamma_{{\rm opt}}\Big|_{{\rm optimal}} & \simeq & \Gamma_{M}(1+2n_{{\rm th}})\sqrt{\frac{4\C}{1+2n_{{\rm th}}}}\nonumber \\
 & \simeq & \Gamma_{M}(1+2n_{{\rm th}})e^{2r}\,.
\end{eqnarray}
One can easily confirm that this is exactly the rate at which the
$\beta$ mode is heated by the mechanical bath (in the large-$r$
limit). We thus see that the optimal coupling condition represents
a simple impedance matching: the rate at which the engineered reservoir
(the cavity) extracts quanta from the $\beta$ mode should match the
rate at which it is {}``heated'' by the intrinsic mechanical dissipation.

\section{Appendix: Squeezed State Purity\label{sec:Appendix:-Squeezed-State}}

In general, the effective thermal occupancy $\neff$ (quantifying
the purity of the mechanical state) is defined by
\begin{equation}
4\left\langle \hat{X}_{1}^{2}\right\rangle \left\langle \hat{X}_{2}^{2}\right\rangle =\left(1+2\neff\right)^{2}\,,\label{eq:def_neff}
\end{equation}
cf. Eq. (10), since $\left\langle \hat{X}_{1}\hat{X}_{2}+\hat{X}_{2}\hat{X}_{1}\right\rangle =0$.
The two variances in terms of the Bogoliubov mode $\be$ read
\begin{align}
2\left\langle \hat{X}_{1}^{2}\right\rangle  & =e^{-2r}\left(1+2\left\langle \bed\be\right\rangle +\left\langle \be\be\right\rangle +\left\langle \bed\bed\right\rangle \right)\nonumber \\
2\left\langle \hat{X}_{2}^{2}\right\rangle  & =e^{2r}\left(1+2\left\langle \bed\be\right\rangle -\left\langle \be\be\right\rangle -\left\langle \bed\bed\right\rangle \right)\,.\label{eq:mech_var_bogol}
\end{align}
As discussed in the main text, optimal squeezing involves a tradeoff
between maximizing the squeeze parameter $r$ (which requires large
$G_{+}/G_{-}$) and maximizing the effective coupling $\mathcal{G}$
to the Bogoliubov mode (which requires small $G_{+}/G_{-}$). The
maximum squeezing at fixed $\C$ thus corresponds to $\beta$ not
being in its vacuum state. Eqs.~(\ref{eq:def_neff}) and (\ref{eq:mech_var_bogol})
thus imply that $\neff\not=0$, cf. Fig.~(2). The optimally squeezed
state is thus in general a thermal squeezed state. 

As is also discussed in the main text, there is a general regime where
one has large optimized squeezing while at the same time having an
\emph{almost }pure state. This occurs for the {}``moderately-strong
coupling'' regime where $\C\gg1$ while at the same time $\mathcal{G}<\kappa$
(no strong coupling hybridization of $\hat{\beta}$ and the cavity).
The latter condition is satisfied for optimized couplings as long
as $\C$ is small enough to satisfy:
\begin{equation}
\C\leq\frac{1}{16}\left(\frac{\kappa}{\Gamma_{M}}\right)^{2}\frac{1}{2\nth+1}\,.\label{eq:strong_coupling_cond-1}
\end{equation}
In this regime, one can achieve optimized steady-state squeezing well
beyond $3\,\mathrm{dB}$ and a low entropy: $\neff\sim\left(\sqrt{2}-1\right)/2\approx0.2$.
(cf.~Fig.~2). 

To see this, we now focus on the limit of large cooperativities $\C=4G_{-}^{2}/\left(\kappa\Gamma_{M}\right)$
while keeping away from the strong coupling regime. Formally, one
can perform this limit by sending $\Gamma_{M}\to0$ (in contrast to
the previous discussions) while keeping all other parameters fixed.
Then, for $\G<\kappa$ and for large $\C$ we find
\begin{equation}
\left(1+2\neff\right)^{2}\approx2\left(1+\frac{4G_{-}^{2}\left(1+2\nth\right)+\kappa^{2}\nth}{\kappa^{2}\sqrt{1+2\nth}}\frac{1}{\sqrt{C}}\right)\to2\label{eq:limit_neff_gamma_zero}
\end{equation}
or $\neff\approx\left(\sqrt{2}-1\right)/2$. Thus, independent of
the choice of parameters, one always finds that the effective thermal
occupancy $\neff\sim\left(\sqrt{2}-1\right)/2$ in the moderately-strong
coupling limit (i.e.~$\C\gg1$ while $\mathcal{G}<\kappa$). Note
that in this regime, the the Bogoliubov mode $\beta$ is characterized
by $\left\langle \bed\be\right\rangle ,\left\langle \be\be\right\rangle \to1/4$.
This follows from the fact that $\left\langle \be\be\right\rangle \to\left\langle \bed\be\right\rangle $
as $\C\to\infty$ and Eqs. (\ref{eq:def_neff})-(\ref{eq:limit_neff_gamma_zero}).

\section{Appendix: Comparison against Measurement-Based Schemes\label{sec:Appendix:-Comparison-against}}

\subsection{Comparison against measurement-based feedback squeezing }

As discussed in the main text, the ability of our scheme to generate
large amounts of stationary quantum squeezing with low entropy indicates
that it outperforms what is possible with a simple coherent parametric
driving (i.e.~spring constant modulation). Here, we also suggest
that it has significant advantages compared to schemes for squeezing
based on a backaction-evading (BAE) single-quadrature measurement
plus feedback \cite{2008_Clerk_BackActionEvasion}. As our scheme
can be viewed as a kind of coherent feedback operation, this apparent
advantage is reminiscent of claims made in Ref.~\cite{2012_HamerlyMabuchi_CoherentFeedback}.
That work also provides specific examples where coherent feedback
control schemes can outperform Gaussian measurement-based schemes

For simplicity, we focus on the regime of large cooperatives $\C=4G_{-}^{2}/\left(\kappa\Gamma_{M}\right)$
where large squeezing is possible both in our scheme and the BAE measurement
scheme. Since the latter scheme was only analyzed in the limit of
no strong coupling effects, we consider the same limit here: we keep
$G_{+},G_{-}\ll\kappa$ while having $\C\rightarrow\infty$ by taking
$\Gamma_{M}\rightarrow0$ . If we simply focus on the maximum possible
squeezing achievable at a fixed cooperativity, there is no fundamental
advantage of our dissipative scheme over the BAE scheme, as both predict
a scaling:
\[
2\left\langle \hat{X}_{1}^{2}\right\rangle \approx\sqrt{\frac{1+2\nth}{\C}}\,.
\]
Of course in practice, achieving this value using BAE measurement
and feedback could be very challenging, as it requires near-ideal
measurements and feedback.

However, the advantage of our coherent feedback scheme (even on an
ideal, fundamental level) becomes apparent when studying the purity
of the generated squeezed state. As already discussed, if we stay
out of the strong coupling regime, the mechanical squeezed state is
almost in a pure state, with the effective number of thermal quanta
$n_{{\rm eff}}$ tending to $\sim0.2$ for large $\C$ as per Eq.~(\ref{eq:limit_neff_gamma_zero}).
In contrast, the BAE measurement plus feedback scheme yields
\[
\left(1+2\neff\right)^{2}=\sqrt{1+2\nth}\sqrt{\C}\to\infty
\]
 in the same limit. Thus, our scheme (an example of coherent feedback)
yields a far more pure state than the measurement-plus-feedback approach.
This represents a significant advantage over the measurement-based
approach.

We note that one could improve the state purity achieved in the BAE
measurement scheme by measuring \textit{both} quadratures of the cavity
output (instead of measuring only the quadrature that contains information
on the coupled mechanical quadrature $\hat{X}_{1}$). One would thus
also learn something about the backaction noise driving the unmeasured
quadrature $\hat{X}_{2}$. This would reduce the conditional variance,
and thus improve the state purity. The analogous situation involving
the dispersive measurement of a qubit is well studied, see e.g. \cite{2013_Hatridge_QuantumBackAction}.

\subsection{Comparison against stroboscopic measurements}

An alternative way of generating mechanical resonator squeezing is
to perform a stroboscopic QND position measurement as suggested in
Ref. \cite{2005_Ruskov_SqueezingMechResByQND_Stroboscopic}. In this
scheme, the measurement rate is modulated in time periodically. The
back-action evading scheme is basically a stroboscopic scheme with
a particular choice for how the measurement rate is modulated (i.e.
sinusoidally). Instead of the cooperativity, the crucial parameter
determining the amount of squeezing generated is now given by the
measurement rate. Based on the analysis of Ref. \cite{2005_Ruskov_SqueezingMechResByQND_Stroboscopic},
the scaling of squeezing and state purity of the stroboscopic measurement
scheme is essentially the same as in the back-action evading scheme.

\subsection{Comparison against pulsed optomechanics schemes}

Let us now compare our dissipative scheme to a pulsed optomechanical
scheme, where large-amplitude pulses of light driving an optomechanical
cavity are used to realize effective strong position measurements,
which, hence, can generate squeezing \cite{2011_Vanner_PulsedQuOptomechanics}.
We note that this scheme requires $\kappa\gg\Omega$. Since this scheme
effectively realizes a strong position measurement, the amount of
squeezing scales with the parameter $\left(\kappa/G\right)^{2}$,
where $G=g_{0}\sqrt{N_{P}}$, and $N_{P}$ is the mean number of photons
per light pulse. This is in marked contrast to our dissipative scheme,
where the cooperativity $\C$ determines the amount of squeezing.
In addition, the pulsed scheme does not generate truly stationary
squeezing, which is again in sharp contrast to our dissipative scheme.
Finally, we note that using the pulsed scheme, it is currently challenging
to get squeezing from this scheme experimentally \cite{2013_Vanner_CoolingByMsmnt}.
However, our scheme should generate large amounts of squeezing even
for state-of-the-art experiments, cf. Fig. 3 of our manuscript.

\section{Appendix: effective Lindblad master equation\label{sec:Appendix:-effective-Lindblad}}

In this section we derive an effective Lindblad master equation which
describes the effects of the cavity (the engineered reservoir) on
the mechanical resonator. Such an approach is common in studies of
reservoir engineering; in contrast, the approach we use in the main
text goes beyond this approximation. We start by considering the Hamiltonian
(6)
\[
\hat{H}=-\hbar\mathcal{G}\dd\be+\text{h.c.}\,,
\]
where $\G^{2}=G_{-}^{2}-G_{+}^{2}$, $\be=\b\cosh r+\bd\sinh r$,
$\cosh r=G_{-}/\G$ and $\sinh r=G_{+}/\G$. Taking the limit of a
large cavity damping rate $\kappa$, one can use standard techniques
\cite{2000_GardinerZoller_QuantumNoise} to eliminate the cavity and
derive a Lindblad-form master equation for the reduced density matrix
of the mechanical resonator $\hat{\rho}.$ This takes the form:
\[
\dot{\hat{\rho}}=\Gamma_{M}\left(\nth+1\right)\mathcal{D}[\b]\hat{\rho}+\Gamma_{M}\nth\mathcal{D}\left[\bd\right]\hat{\rho}+\Gamma_{\text{opt}}\mathcal{D}\left[\be\right]\hat{\rho}\,,
\]
where $\mathcal{D}\left[\hat{A}\right]=\hat{A}\hat{\rho}\hat{A}^{\dagger}-\hat{A}^{\dagger}\hat{A}\hat{\rho}/2-\hat{\rho}\hat{A}^{\dagger}\hat{A}/2$.
Expressing the Bogoliubov mode in terms of the original operators
$\b$ and $\bd$ yields:
\begin{align}
\dot{\hat{\rho}} & =\left[\Gamma_{M}\left(\nth+1\right)+\Gamma_{\text{opt}}\cosh^{2}r\right]\mathcal{D}[\b]\hat{\rho}\nonumber \\
 & +\left[\Gamma_{M}\nth+\Gamma_{\text{opt}}\sinh^{2}r\right]\mathcal{D}\left[\bd\right]\hat{\rho}\nonumber \\
 & +\Gamma_{\text{opt}}\cosh r\sinh r\,\mathcal{D}_{S}\left[\hat{b}\right]\hat{\rho}\nonumber \\
 & +\Gamma_{\text{opt}}\cosh r\sinh r\,\mathcal{D}_{S}\left[\bd\right]\hat{\rho}\label{eq:eff_LME_mechanics}
\end{align}
where $\Gamma_{\text{opt}}=4\G^{2}/\kappa$ and  $\mathcal{D}_{S}\left[\hat{b}\right]\hat{\rho}=\b\hat{\rho}\b-\b\b\hat{\rho}/2-\hat{\rho}\b\b/2$.
The last two terms on the RHS of this equation do not conserve the
number of mechanical quanta, and are directly responsible for the
generation of squeezing. This Lindblad master equation is similar
to the one discussed in the context of dissipative preparation of
spin squeezed atomic ensembles \cite{2013_DallaTorre_SteadyStateSqueezingSpinEnsemble}.

Using Eq.~(\ref{eq:eff_LME_mechanics}), we can again calculate the
optimal value of $G_{+}$ which maximizes the squeezing (with other
parameters and $G_{-}$ held fixed). We find that this approach leads
to the same expression which we have already found using the full
theory, Eq.~(8) in the main text. For this optimized coupling and
in the limit of large cooperativity $\C$, the Lindblad approach predicts
\[
2\left\langle \hat{X}_{1}^{2}\right\rangle \approx\sqrt{\frac{1+2\nth}{\C}}.
\]
Comparing against Eq.~(9) in the main text, we see that the Lindblad
approach misses the saturation of squeezing to $\Gamma_{M}\left(1+2\nth\right)/\kappa$
in the large $\C$ limit. The approximations used to derive the Lindblad
master equation tacitly assume $\kappa\to\infty$, and thus neglect
the finite rate at which the cavity is able to expel energy extracted
from the mechanical resonator. 

When focussing on the purity of the squeezed state, we find that the
Lindblad master equation predicts
\[
\left(1+2\neff\right)^{2}\approx2+\frac{2\nth}{\sqrt{2\nth+1}}\frac{1}{\sqrt{C}}\,.
\]
This is in strong contrast to the prediction $\left(1+2\neff\right)^{2}\sim\sqrt{\C}$
of the full theory. This is, because the Lindblad master equation
cannot capture strong coupling effects.

\section{Appendix: Cavity output spectra and Squeezing Detection\label{sec:Appendix:-Cavity-output}}

\begin{figure}
\includegraphics{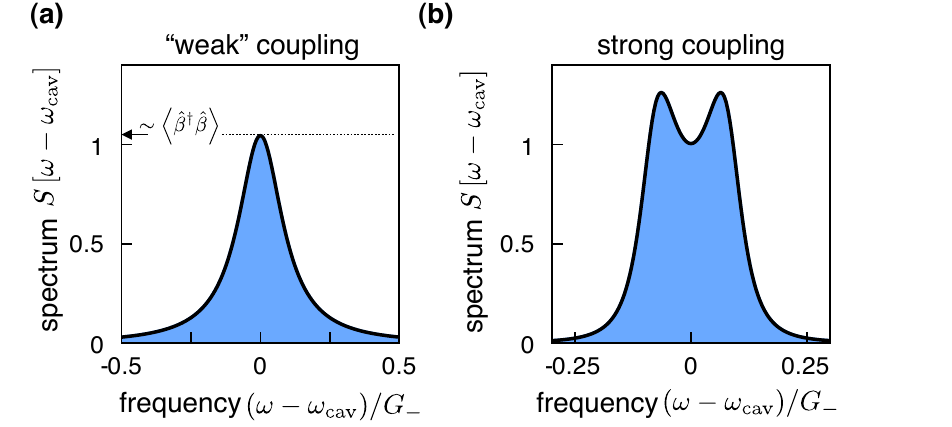}

\caption{Cavity output spectra. (a) Cavity output spectrum for weak coupling.
The spectrum at $\omega-\omega_{\text{cav}}$ as well as the area
of the spectrum is directly proportional to the occupancy $\left\langle \bed\be\right\rangle $.
(b) If we increase the cooperativity, we observe a normal mode splitting.
This is a signature of the {}``strong coupling regime'', where the
Bogoliubov mode and the photons hybridize. {[}Parameters: (a) $\Gamma_{M}/\kappa=10^{-4}$,
$\nth=10$ and $\C=10^{4}$. (b) Same as (a) but $\C=10^{6}${]}.}

\label{fig:cavity_output_spectra}
\end{figure}
Let us consider the cavity output spectrum 
\[
S\left[\omega\right]=\int\mathrm{d}t\, e^{\i\omega t}\left\langle \delta\hat{a}_{\text{out}}^{\dagger}\left(t\right)\delta\hat{a}_{\text{out}}(0)\right\rangle \,,
\]
where $\delta\hat{a}_{\text{out}}=\hat{a}_{\text{out}}-\left\langle \hat{a}_{\text{out}}\right\rangle $
and $\hat{a}_{\text{out}}+\hat{a}_{\text{in}}=\sqrt{\kappa}\hat{a}$
for an ideal, single-sided cavity \cite{2008_ClerkDevoretGirvinFMSchoelkopf_RMP}.
We find that
\[
S\left[\omega-\omega_{\text{cav}}\right]=\frac{16\kappa\Gamma_{M}\left[G_{+}^{2}\left(\nth+1\right)+G_{-}^{2}\nth\right]}{\left|N[\omega]\right|^{2}}\,,
\]
where
\[
N\left[\omega\right]=4\G^{2}+\left(\Gamma_{M}-2\i\omega\right)\left(\kappa-2\i\omega\right)\,.
\]
As shown in Fig. \ref{fig:cavity_output_spectra}, the coupling to
the $\beta$ mode gives rise to weight in the output spectrum near
the cavity resonance frequency. For weak coupling ($\mathcal{G}<\kappa$)
, one has a simple Lorentzian peak, whereas for a strong coupling
($\mathcal{G}>\kappa$) a double-peak structure emerges. The condition
for this strong coupling to occur was given in Eq.~(\ref{eq:strong_coupling_cond-1}).

As discussed in the main text (cf.~Eq.~(12)), one can detect the
squeezing of the mechanical resonator by first measuring $\langle\hat{\beta}^{\dagger}\hat{\beta}\rangle$
from the integrated output spectrum, and then using this to bound
the variance of the $\hat{X}_{1}$ quadrature. The general expression
for the $\Xone$ variance in terms of the Bogoliubov mode $\be$ is
given in Eq.~(\ref{eq:mech_var_bogol}). Since in general 
\[
\left|\left\langle \be\be\right\rangle \right|\le\left\langle \bed\be\right\rangle +\frac{1}{2}
\]
(which can be shown by using the Cauchy-Schwarz inequality), one finds
a general upper bound for the squeezing
\[
2\left\langle \hat{X}_{1}^{2}\right\rangle \le2e^{-2r}\left[1+2\left\langle \bed\be\right\rangle \right]\,.
\]
In the limit of large cooperativity, and, hence, large $r$, one can
also find a lower bound by making use of the theoretical predictions
{[}for $\kappa/\Omega=0${]}. For any value of the squeezing parameter
$r$ we find that $\left\langle \be\be\right\rangle \sim\left\langle \bed\be\right\rangle $.
In the limit of large cooperativities {[}and, hence, large $r${]},
\begin{equation}
\left\langle \be\be\right\rangle \approx\left[1+\left(4\frac{\nth+1}{2\nth+1}-2\right)e^{-2r}\right]\left\langle \bed\be\right\rangle \,.\label{eq:bebe_sim_bedbe}
\end{equation}
Thus,
\[
2\left\langle \hat{X}_{1}^{2}\right\rangle \approx e^{-2r}\left[1+4\zeta\left\langle \bed\be\right\rangle \right]
\]
where 
\begin{equation}
\zeta=1+\left(2\frac{1+\nth}{1+2\nth}-1\right)e^{-2r}\ge1\,.\label{eq:estimate_zeta}
\end{equation}
Using this estimate, we finally find the lower bound
\[
2\left\langle \hat{X}_{1}^{2}\right\rangle \ge e^{-2r}\left(1+4\left\langle \bed\be\right\rangle \right)\,.
\]

\section{Appendix: Effects of finite sideband parameters\label{sec:Appendix:-Effects-of}}

\begin{figure}[t]
\includegraphics{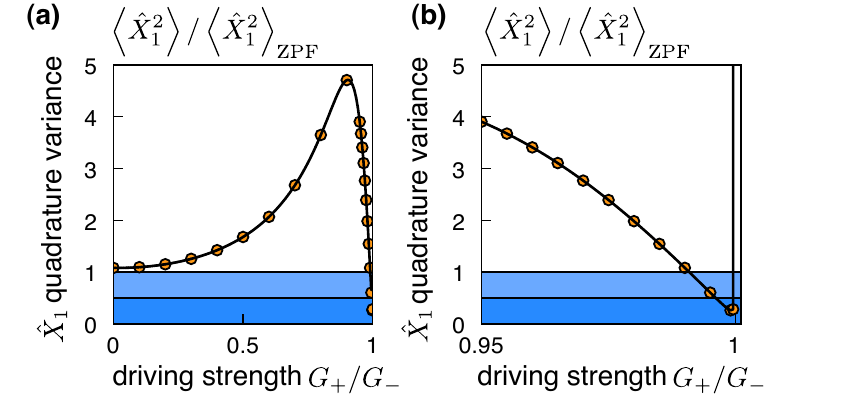}

\caption{(a) Quadrature variance $\left\langle \hat{X}_{1}^{2}\right\rangle $
as a function of the blue driving strength $G_{+}/G_{-}$ including
finite sideband parameter effects. The black curve shows the analytical
result whereas the circles represent findings due to a numerical simulation
of the full Hamiltonian (3). (b) Zoom of (a). We see that the mechanical
resonator can be squeezed beyond the $3\,\mathrm{dB}$ limit. {[}Parameters:
$\Gamma_{M}/\kappa=10^{-4}$, $\kappa/\Omega=1/50$ , $\nth=100$
and $\C=5\cdot10^{6}$ {]}.}

\label{fig:X1_sq_vs_gp_incl_CR}
\end{figure}

In order to consider effects due to finite sideband parameters, we
perturbatively solve the quantum Langevin equations for the cavity
and mechanical resonator operators by using the full, time-dependent
Hamiltonian (3), keeping the leading corrections in $\kappa/\Omega$.
We also compare against a full numerical solution of the equations.

\begin{figure}
\includegraphics{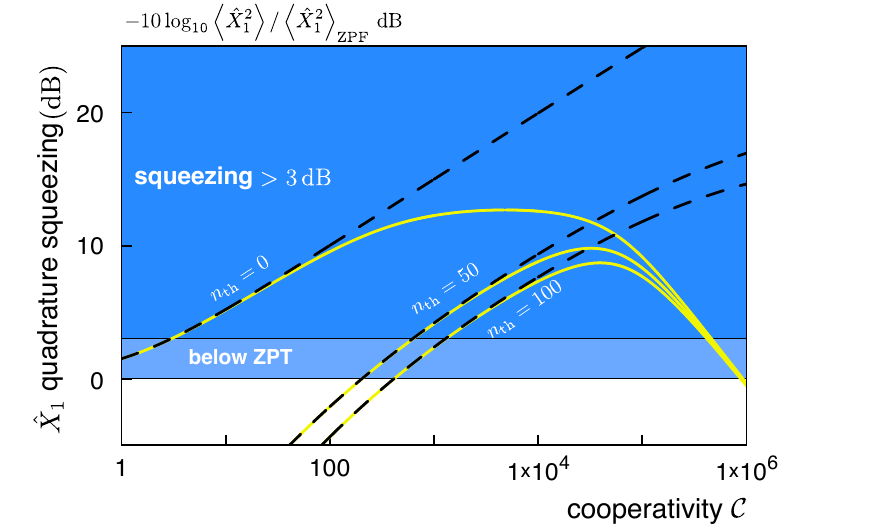}

\caption{Squeezing as a function of the cooperativity for optimized driving
strengths where realistic experimental parameters are assumed \cite{2011_Chan_LaserCoolingNanomechOscillator}.
The yellow line and the white dashed lines show the theory \textit{including}
effects due to a finite sideband parameter. The black, dashed line
shows the expectation for $\kappa/\Omega=0$. Thus, squeezing beyond
$3\,\mathrm{dB}$ is expected for state-of-the-art experiments. {[}Parameters:
$\Gamma_{M}/\kappa=10^{-4}$, $\kappa/\Omega=1/10${]}.}

\label{Figure_6}
\end{figure}

In Fig.~\ref{fig:X1_sq_vs_gp_incl_CR} the quadrature variance $\left\langle \hat{X}_{1}^{2}\right\rangle $
is shown as a function of the blue laser driving strength $G_{+}/G_{-}$
including bad cavity effects. The black curve depicts our analytical
perturbative expression (which is too lengthy to be reported here),
whereas the orange circles show the result of a numerical simulation
of (3). Note that $\left\langle \hat{X}_{1}^{2}\right\rangle $ is
strongly non-monotonic. We find that a unique optimum of the driving
strength $G_{+}/G_{-}$ maximizing steady-state squeezing still exists
(cf.~Fig.~\ref{fig:X1_sq_vs_gp_incl_CR}(b)). A plot showing the
maximized squeezing as a function of the cooperativity $\C$ for fixed
sideband parameter $\kappa/\Omega$ and decay rates $\Gamma_{M}/\kappa$
is shown in Fig. 3, where parameters of a state-of-the-art experiment
\cite{2011_Teufel_SidebandCooling_Nature} have been assumed. Fig.~\ref{Figure_6}
also shows maximized squeezing as a function of $\C$ for experimental
parameters of state-of-the-art photonic crystal experiments \cite{2011_Chan_LaserCoolingNanomechOscillator}.

Let us now discuss the influence of the counter-rotating terms in
more detail. For small cooperativities the effect of the counter-rotating
terms is small. As we increase the cooperativity, the squeezing parameter
$r$ also increases (cf.~Eq.~(8) for the case $\kappa/\Omega=0$),
such that the counter-rotating terms become more and more important.
Since $r\approx\ln\left[4\C/\left(1+2\nth\right)\right]/4$ for $\kappa/\Omega=0$
and large $\C$, we find that the smaller $\nth$, the earlier these
corrections become important. When increasing the cooperativity further,
maximum squeezing is assumed first after which squeezing gets lost
again. 

\section{Appendix: Avoiding the Parametric Instability\label{sec:Appendix:-Avoiding-the}}

It turns out that due to the two-tone driving, the radiation pressure
force $F\propto\left|\bar{a}_{+}e^{-\i\omega_{+}t}+\bar{a}_{-}e^{-\i\omega_{-}t}\right|^{2}$
oscillates at twice the mechanical frequency, since $\omega_{\pm}=\omega_{\text{cav}}\pm\Omega$.
In an experiment, these oscillations can yield parametric instabilities,
if the mechanical resonator frequency $\Omega$ is strongly temperature
dependent \cite{2012_SuhBAEParampInstability}. To suppress this instability,
one can add a third driving tone to cancel the oscillations of the
radiation pressure force at $2\Omega$ \cite{2013_Steinke_BackacktionEvadingMsrmt}.
In the following, we show that the influence of this third driving
tone on the generation of steady-state squeezing is small for typical
experimental parameters.

The third driving tone can be included in our theory by adding the
driving term
\[
\hat{H}_{\text{drive,add}}=\hbar\alpha_{3}\left(e^{-\i\omega_{3}t+\i\varphi}\ad+\text{h.c.}\right)
\]
to the Hamiltonian (1). One then finds that the radiation pressure
force reads
\[
F\propto\left|\bar{a}_{+}e^{-\i\omega_{+}t}+\bar{a}_{-}e^{-\i\omega_{-}t}+\bar{a}_{3}e^{-i\omega_{3}t+\i\varphi}\right|^{2}\,.
\]
When choosing $\omega_{3}=\omega_{\text{cav}}-3\Omega$ \cite{2013_Steinke_BackacktionEvadingMsrmt}
we find that the component of $F$ oscillating at $2\Omega$ vanishes
if $\varphi=\pi$ and $\bar{a}_{3}=\bar{a}_{+}$. Linearizing the
resulting Hamiltonian again and going into an interaction picture
with respect to the free cavity and mechanical resonator Hamiltonian,
we find that the third drive tone gives rise to an additional term
\begin{equation}
\hat{H}_{\text{new}}=\hbar G_{3}e^{3\i\Omega t}\left[\b e^{-\i\Omega t}+\bd e^{\i\Omega t}\right]\dd+\text{h.c.}\label{eq:H_new_due_to_third_tone}
\end{equation}
to the linearized Hamiltonian (3), where $G_{3}=g_{0}\bar{a}_{3}=G_{+}$. 

\begin{figure}
\begin{centering}
\includegraphics{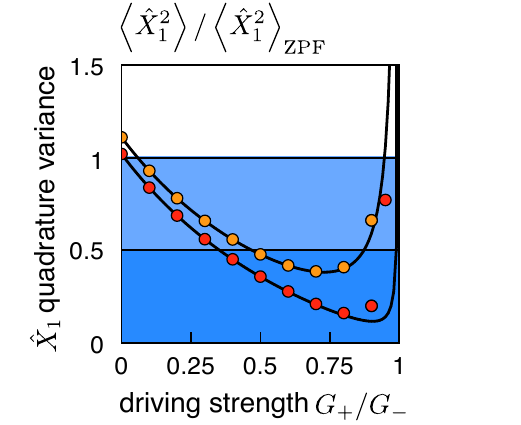}
\par\end{centering}

\caption{Quadrature variance $\left\langle \hat{X}_{1}^{2}\right\rangle $
as a function of the blue driving strength $G_{+}/G_{-}$ including
finite sideband parameter effects and effects of a third tone, cf.
Eq. (\ref{eq:H_new_due_to_third_tone}). The black curve $\left\langle \hat{X}_{1}^{2}\right\rangle $
including effects of finite sideband parameters but ignores the third
tone, i.e. $G_{3}=0$. The circles represent results of a numerical
simulation of the full Hamiltonian including finite sideband effects
and effects due to the third tone. {[}Parameters: $\Gamma_{M}/\kappa=10^{-4}$,
$\kappa/\Omega=1/50$, $\nth=50$ and $\C=10^{3}$ {[}upper curve{]}
and $\C=10^{4}$ {[}lower curve{]}, respectively{]}.}

\label{FIG_7}
\end{figure}

The impact of this additional counter rotating term on the generation
of steady-state squeezing is shown in Fig. \ref{FIG_7}. In this figure,
the quadrature variance $\left\langle \hat{X}_{1}^{2}\right\rangle $
is shown as a function of the blue laser driving strength $G_{+}/G_{-}$
including all counter rotating terms. For a medium cooperativity,
the influence of the third tone is negligible. For a larger cooperativity,
deviations become visible as $G_{+}\to G_{-}$. However, the minimum
value of $\left\langle \hat{X}_{1}^{2}\right\rangle $ is changed
little and lies still well beyond $3\,\mathrm{dB}$. 

Let us now briefly discuss why the deviations become apparent as we
increase the cooperativity and as we approach $G_{+}\to G_{-}$. An
increase of the cooperativity $\C$ leads to an increase of the squeezing
parameter $r$, such that the influence of counter rotating terms
becomes larger, cf. the discussion in the previous section. Thus,
the influence of the additional, counter rotating terms (\ref{eq:H_new_due_to_third_tone})
increases with increasing $\C$. Additionally, to cancel the unwanted
frequency component $2\Omega$ of the radiation pressure force, we
have to choose $G_{3}=G_{+}$. Thus, as we increase $G_{+}$, the
magnitude of the additional Hamiltonian (\ref{eq:H_new_due_to_third_tone})
also increases, leading to a larger perturbation of the steady-state
quadrature variance.

Thus, to conclude, we can avoid the parametric instability by adding
a third driving tone while still generating squeezing well beyond
$3\,\mathrm{dB}$.

\bibliographystyle{apsrev4-1}
\bibliography{Optomechanics}

\end{document}